\documentstyle[aps,prl,multicol]{revtex}
\title{Universality of Decoherence}
\author{Daniel Braun, Fritz Haake, Walter T. Strunz}
\address{Fachbereich Physik, Universit\"at-GH Essen\\
45117 Essen\\ Germany}

\begin{document}
\maketitle
\begin{abstract}
We consider environment induced decoherence of quantum superpositions to
mixtures in the limit in which that process is much faster than
any competing one generated by the Hamiltonian $H_{\rm sys}$ of the
isolated system. While the golden rule then does not apply we can discard
$H_{\rm sys}$. By allowing for simultaneous couplings to different
reservoirs, 
we reveal decoherence as a universal short-time phenomenon independent
of the character of the system as well as the bath and of the basis the
superimposed states are taken from. We discuss consequences for the
classical behavior of the macroworld and quantum measurement: For the
decoherence of superpositions of macroscopically distinct states the system
Hamiltonian is always negligible.  
\end{abstract}
\begin{multicols}{2}
\section*{\strut}
Superpositions of quantum states give rise to interference effects which
are, however, more and more difficult to observe as the size of the
system is increased and the  superimposed states are made more distinct.
While quantum interferences are ubiquitous in the microworld,
none have been seen for macroscopic bodies. A parameter
controlling the discernability of interference fringes is the ratio of
the de Broglie wavelength $\lambda$ of a particle to a typical linear
dimension $\Delta$ of the spatial structure used to construct
superpositions of different partial particle waves. When that
parameter is of order unity, interference is easily measurable; but upon
decreasing $\lambda/\Delta$, either by letting $\Delta$ grow or
using ever more massive particles and thus ever smaller $\lambda$,
wave effects become elusive and eventually escape current
detection techniques. As a concrete example, we may think of double-slit
experiments (with $\Delta$ the slit distance) for which the increasing
difficulty in question concerns the resolution of angular apertures of
diffraction structures of order $\lambda/\Delta$.

It is nowadays widely accepted that an even more important reason for the
notorious absence of quantum superpositions from the macroscopic world
lies in environment imposed decoherence \cite{Zurek,Zeh}, and that phenomenon is our concern here. Decoherence is, for
microscopic bodies, just a facet of dissipation caused by
interactions with many-freedom surroundings. Spontaneous emission
of light by an atom is such a dissipative process, with the
electromagnetic field acting as a weakly coupled environment. Inasmuch
as radiatively coupled states of an atom are only ``microscopically''
distinct it would appear overly pedantic and even somewhat misleading to
attach the fashionable label decoherence to polarization decay while reserving
dissipation to the exponential decrease of the population of the
excited state. But ``larger'' systems under the influence of
environments do invite such different names for the following reason.
If two sufficiently distinct states $|s\rangle,|s'\rangle$ are brought to
an initial superposition $|\,\rangle=c|s\rangle+c'|s'\rangle$, the
density operator $\rho(t)$ starts out as the projector
$\rho(0)=|\,\rangle\langle\,|$ and then, for suitable coupling to the
environment (see below), decoheres to the mixture
$|c|^2|s\rangle\langle s|+|c'|^2|s'\rangle\langle s'|$, with the weights
$|c|^2,|c'|^2$ still as in the initial superposition, on a time scale
$\tau_{\rm dec}$ while the subsequent relaxation of that mixture towards
an eventually stationary $\rho(\infty)$ has a much longer characteristic
time $\tau_{\rm diss}$. The time scale ratio $\tau_{\rm dec}/\tau_{\rm
diss}$ becomes the smaller the more distinct the two component states
are. If that distinction can be measured by a length
$\Delta\propto|s-s'|$ the
time scale ratio is again determined by the
ratio of the de Broglie wavelength to $\Delta$, typically as
\begin{equation}
  \tau_{\rm dec}/\tau_{\rm diss}\sim
  \left(\lambda/\Delta\right)^2 \,.
  \label{1}
\end{equation}
The quadratic rather than just linear dependence of the ``acceleration
of decoherence against dissipative equilibration of probabilities'' on
the ratio $\lambda/\Delta$ suggests that
decoherence gives rise to classical behavior of
the macroworld. Indeed, for macroscopic bodies and macroscopic values
of $\Delta$ the acceleration factor is typically so small that
decoherence appears as instantaneous while dissipation for
classically meaningful quantities may not at all be noticeable.

The acceleration factor in question has been studied in recent
experiments \cite{Zeili,Welschi,Winey}. Arndt et al.~in Vienna
\cite{Zeili} observed multiple-slit diffraction of the largest objects thus far, ${\rm C}_{60}$
molecules. No decohering influence of the environment was effective, simply because $\tau_{\rm diss}$ and the relevant thermal de Broglie length were sufficiently large. Experiments 
at the ENS in Paris \cite{Welschi} involved
superpositions of  coherent states of a microwave cavity mode. Even
though the cavity was of high quality ($\tau_{\rm diss}=160\mu
{\rm s}$) the acceleration factor was
controlled between, roughly, 1 and 10. Finally, a NIST group
\cite{Winey} worked with superpositions of translational-motion states
of single $^9{\rm Be}^+$ atoms in Paul traps. Here, the acceleration
factor was steered through the range $1\ldots 25$ and the environment
engineered so as to even vary $\tau_{\rm diss}$.

On the theoretical side the experiments mentioned are well understood.
In all cases the decoherence time is not smaller than the time scale of
dissipation by many orders of magnitude; it is in fact still larger than
the characteristic times of the free motion of the system in the absence
of any environment. In that limit, both dissipation and its companion
decoherence can be treated by Fermi's golden rule or fancied-up variants
thereof like master equations. The golden rule involves a certain
long-time limit: It cannot yield time independent transition rates (or
Markovian master equations) before the time elapsed since the
preparation of the initial state exceeds the basic periods $\tau_{\rm
sys}$ of the isolated system. Indeed, inasmuch as it explicitly requires
energy conservation for the exchange of free-bath and free-system
quanta, it presupposes such large times for resonance to become
effective.

The golden rule cannot be trusted when it predicts a
decoherence time $\tau_{\rm dec}$ smaller than $\tau_{\rm sys}$. It
therefore does not help to explain why the macroworld behaves
classically. When nevertheless holding decoherence
responsible here we mostly rely on an exactly solvable model, a harmonic
oscillator harmonically coupled to a bath itself consisting of
harmonic oscillators \cite{harmonicoscillators}. The acceleration factor
$(\lambda/\Delta)^2$ is there found in effect already for arbitrarily
small times. Invaluable as the oscillator model is for rigorously
revealing decoherence, we would like to prove a purportedly universal
phenomenon to emerge universally, rather than only for a very special
model.

The clue to progress lies in the fact that the Hamiltonian $H=H_{\rm
sys}+H_{\rm bath}+H_{\rm int}$ of the embedding of a system in an
environment (alias bath) can be simplified for times much smaller than
the characteristic times $\tau_{\rm sys}$ of $H_{\rm sys}$: We can
altogether neglect any motion the isolated system would perform,
i.~e.~discard $H_{\rm sys}$. For the structure of the interaction
Hamiltonian we do not have much of a choice. Introducing a coupling
agent each for the system, $S$, and the bath, $B$, we may write
\begin{equation}
  H_{\rm int}=SB.
  \label{2}
\end{equation}
It follows that the coupling agent $S$ becomes conserved and only
plays the role of a fixed parameter. Introducing eigenvectors and eigenvalues of $S$ as
$S|s\rangle =s|s\rangle$ we consider the matrix element
$\langle s|W|s'\rangle$ of the joint density operator of our compound
which still is a density operator for the bath. We shall eventually be
interested only in the reduced density matrix  $\langle
s|\rho|s'\rangle={\rm Tr_{bath}}\langle s|W|s'\rangle$ and intend to
rigorously reveal decoherence as a universal short-time phenomenon by
showing
\begin{equation}
  \langle s|\rho(t)|s'\rangle={\rm e}^{-(s-s')^2f(t)+{\rm
  i}(s^2-s'^2)\,\varphi(t)}\,\langle s|\rho(0)|s'\rangle\,,
  \label{3}
\end{equation}
with functions $f(t)\geq0$ and  $\varphi(t)$ to be determined.

For the next step we momentarily model the bath as a collection of
oscillators and specify
\begin{equation}
  H_{\rm bath}=\sum_{i=1}^N\Big(\frac{1}{2m}\hat{p}_i^2+\frac{1}{2}m\omega_i^2\hat{q}_i^2\Big)\,,\;
  H_{\rm int}=S\sum_{i=1}^Ng_i\hat{q}_i \,.
  \label{4}
\end{equation}
To write the Liouville-von Neumann equation for the joint density
operator $W$ it is convenient to stick to the $S$-representation and
employ the Wigner function with respect to the bath oscillators.
Denoting by $W(s,s',p,q,t)$ that hybrid representative we get the
reduced density matrix by integrating over the $(2N)$-dimensional phase
space of the bath, $\langle s|\rho(t)|s'\rangle=\int
d^Npd^NqW(s,s',p,q,t)$. The evolution equation $\dot{W}=LW$ has the
generator
\begin{displaymath}
  L\!=\!\sum_i\!\!\left[\!\frac{\partial}{\partial p_i}m\omega_i^2q_i-
  \frac{\partial}{\partial q_i}\frac{p_i}{m}
  -g_i\!\Big[\frac{{\rm i}}{\hbar}(s-s')q_i-\frac{s+s'}{2}\frac{\partial}{\partial
  p_i}\Big]\!\!\right]\!.
\end{displaymath}
The purely parametric role of the eigenvalues $s,s'$ of the system
coupling agent is manifest in the generator $L$.
We could proceed to solving the foregoing first-order differential
equation for $W(s,s',p,q,t)$. It is
more convenient to directly go for the time evolution of the reduced
density matrix. To that end we may assume initial
statistical independence of system and bath, $W(s,s',p,q,0)=\langle
s|\rho(0)|s'\rangle \times W_{\rm bath}(p,q,0)$. Without loss of
generality we momentarily assume the initial bath distribution sharp,
$W_{\rm bath}(p,q,0)=\prod_i\delta(p_i-p_{i0})\delta(q_i-q_{i0})$, since
we may later average with whatever weight we please. A reduced
time evolution operator can be introduced as
$U(t)=\int\,d^Npd^Nq \,{\rm e}^{Lt}\,W_{\rm bath}(p,q,0)$.
We readily check
\begin{equation}
  \dot{U}=-\frac{{\rm i}}{\hbar}(s-s')\sum_ig_i\int\!d^N\!pd^N\!q\,q_i\,{\rm
  e}^{Lt}\,W_{\rm bath}(p,q,0)
  \label{6}
\end{equation}
and then, using the commutator $[\frac{\partial}{\partial
q_i},q_j]=\delta_{ij}$, shift the factor $q_i$ in the integrand to the
right of the exponential ${\rm e}^{Lt}$, to get a reduced generator
$l(t)=\dot{U}(t)U(t)^{-1}$ as
\begin{eqnarray}
  l(t)=\frac{{\rm i}}{\hbar}\sum_i\Big[&&(s^2-s'^2)\frac{g_i^2}{2m\omega_i^2}(1-\cos\omega_it)
        \nonumber\\
       &&-(s-s')g_i\big(q_{i0}\cos\omega_i t
       +\frac{p_{i0}}{m\omega_i}\sin\omega_i t\big)\Big] \,.
  \label{7}
\end{eqnarray}
Due to the parametric role of the eigenvalues $s,s'$ we
here do not confront a differential operator and get
the density matrix as $\langle
s|\rho(t)|s'\rangle=\exp\{\int_0^t\,dt'l(t')\}\,\langle
s|\rho(0)|s'\rangle\,;$
this contains the initial coordinates $q_{i0}$ and momenta
$p_{i0}$ in the exponent. Now we invoke a thermal
bath and average as
$\overline{{\rm e}^{{\rm i}aq_i}}={\rm e}^{-a^2\hbar/4m\omega_i\tanh(\beta\hbar\omega_i/2)}$
and $\overline{{\rm e}^{{\rm i}bp_i}}={\rm
e}^{-b^2m\hbar\omega_i/4\tanh(\beta\hbar\omega_i/2)}$\,.
The result (\ref{3}) is so reached with
\begin{eqnarray}
        f(t)\!&=&\!\sum_i\!\frac{g_i^2(1\!+\!2\overline{n}_i)}{2m\hbar\omega_i^3}
        \big(1 \!-\!\cos\omega_it\big)={\rm Re}\frac{1}{\hbar^2}
  \!\int_0^t\!\!\!dssC(t-s)\,,\nonumber\\
  \varphi(t)\!&=&\!\!\sum_i\!\frac{g_i^2}{2m\hbar\omega_i^2}\big(t\!-
  \!\frac{\sin\omega_it}{\omega_i}\big)={\rm Im}\frac{1}{\hbar^2}
  \!\int_0^t\!\!\!dssC(t-s)\,,
  \label{9}
\end{eqnarray}
where $\overline{n}_i=(
{\rm e}^{\beta\hbar\omega_i}-1)^{-1}$ is the thermal number of quanta and
$C(t)=\langle B(t)B(0)\rangle$ the thermal autocorrelation function of
the bath coupling agent. The function $f(t)$, which determines the
decoherence as thermally enhanced by the factor $1+2\overline{n}_i$,
begins quadratically in $t$; for larger times it approaches $f(t)\to
\gamma t$ with $\gamma={\rm Re}\,\hbar^{-2}\int_0^\infty\!dtC(t)$, provided $C(t)$ falls off faster than $t^{-2}$. The proportionality of $f(t)$ to $\hbar^{-1}$ signals a quantum scale $\lambda^2$ of reference for $(s-s')^2$. The phase $\varphi(t)$ begins $\propto t^3$
and is temperature independent. We should appreciate the dramatic
difference of the decoherence function $f(t)$ from its golden-rule
counterpart. If the system were itself an harmonic oscillator with
frequency $\Omega$, mass $M$, and displacement $S$ the golden-rule would yield
$f_{\rm GR}(t)=\gamma^{\rm GR}t$ with $\gamma^{\rm GR}=(1+2\overline{n}(\Omega))(2\hbar)^{-2}{\rm Re}\int_0^\infty\!dt{\rm e}^{{\rm i}\Omega t}\langle [B(t),B(0)]\rangle$.
Most importantly, our decoherence function in (\ref{3}) describes
accelerated decoherence for whatever system with whatever coupling agent
$S$, provided only decoherence is fast in the sense $\tau_{\rm
dec}\ll\tau_{\rm sys}$ which will always be the case for sufficiently
distinct $s$ and $s'$.

We have not gone more than half way towards our goal yet. The
decoherence shown by the foregoing reasoning is a privilege of
superpositions of eigenstates of the system coupling agent;
superpositions of eigenstates of other system observables not commuting
with $S$ do not lose their relative quantum phases
any faster than probabilities change. But on the other hand no
privileged representations are known in the macroworld; no quantum
coherence has ever been seen between any pair of macroscopically distinct
states.

For fast decoherence of macroscopic superpositions to take place without
distinction of special observables or states, a variety of environmental
influences would have to be at work. Rather than
privileging a single observable $S$ as the one and only coupling agent
to contact one and only one reservoir it seems necessarry to
account for several non-commuting system observables as coupling agents
toward several reservoirs.

To model the situation just sketched we accompany the single agent $S$ by
a canonically conjugate partner $R$ with $[R,S]=\hbar/{\rm i}$. More general
models would involve a larger set of non-commuting system observables
but would not lead to conclusions qualitatively different from the ones
to be discussed here. Still interested in times smaller than $\tau_{\rm
sys}$ we need not worry about a system Hamiltonian $H_{\rm sys}$ and
generalize (\ref{4}) to the double-bath Hamiltonian
\begin{eqnarray*}
  H_{\rm bath}&=&\!\sum_{i}\!\Big[\frac{1}{2m}\hat{p}_i^2\!+\!\frac{1}{2}m\omega_i^2\hat{q}_i^2\Big]
 \!+\!\!\sum_{i}\!\Big[\frac{1}{2M}\hat{P}_i^2\!+\!\frac{1}{2}M\Omega_i^2\hat{Q}_i^2\Big]
              ,\nonumber\\
  H_{\rm int}&=&S\sum_{i}g_i\hat{q}_i+R\sum_{i}G_i\hat{Q}_i \,.
\end{eqnarray*}
Neither $S$ nor its conjugate
partner $R$ now play a merely parametric role. Even though
the neglect of $H_{\rm sys}$ still forbids the appearance of any nonlinearity
and guarantees explicit tractability with the same strategy as
above, the resulting expressions are so unwieldy to not warrant
full display. The generator $L$ gets an addition
differing from the previous term only by
$\{p_i,q_i,\omega_i,m\}\to\{P_i,Q_i,\Omega_i,M\}$ and
$s\to\frac{\hbar}{{\rm i}}\frac{\partial}{\partial s}$,
$s'\to{\rm i}\hbar\frac{\partial}{\partial s'}$, the latter two
replacements familiar for a momentum in the position representation.
The corresponding addition arises for the time rate of change of the
reduced evolution operator $\dot{U}(t)$ in (\ref{6}) but from that point
on an avalanche of complications is set loose: When scrutinizing
${\rm e}^{-Lt}q_i{\rm e}^{Lt}$ and ${\rm e}^{-Lt}Q_i{\rm e}^{Lt}$ we
find all variables $\{q_i,p_i,Q_i,P_i,s,\frac{\partial}{\partial
s}\}$ coupled. The analogue of the reduced generator
(\ref{7}) is bulky and, worse, contains non-commuting pieces and
thus gives rise to a yet bulkier time ordered exponential
$(\exp\int_0^tdt'l(t'))_+$. Fortunately, to within
corrections of order $t^3$ we can drop all
non-commuting terms in the new $l(t)$ and arrive at the
reduced density matrix
\begin{eqnarray}
   \langle s|\rho(t)|s'\rangle={\rm e}^{-(s-s')^2f(t)}\,{\rm e}^{\hbar^2
    ({\textstyle \frac{\partial}{\partial s}}
   +{\textstyle \frac{\partial}{\partial s'}})^2F(t)}
   \,\langle s|\rho(0)|s'\rangle
\label{11}
\end{eqnarray}
where $f(t)$ is given by (\ref{9}) and $F(t)$ likewise except for
$(m,\omega_i,g_i)\to(M,\Omega_i,G_i)$. To make peace with the absence of
a phase factor from (\ref{11}) it is well to recall from (\ref{9}) that
$\varphi(t)\propto t^3$ for small times. Upon Fourier transforming to
the momentum representation we see that the second exponential in
(\ref{11}) entails accelerated decay for coherences with respect to
eigenstates of the momentum $R$ just as the first factor does for
the eigenstates of $S$.

As expected, by introducing different reservoirs coupling to
non-commuting observables we break the privilege of a single
representation. Non-commuting agents contacting different
reservoirs occur for a body probing an electric field through its charge
or electric dipole moment and a magnetic field with its magnetic moment.
That simple example indicates that for a macroscopic body the exclusive
action of a single reservoir may be as unrealistic a fiction
as complete isolation.

Our discussion is of relevance for the quantum measurement problem. It
had long been considered a puzzle how a microscopic object prepared in a
superposition of, say, two eigenstates of an observable to be measured
can, through unitary evolution of its composition with a macroscopic
pointer, cause that pointer to reach one of two distinct
positions in each run of the measurement, with those 
positions uniquely related to the two eigenvalues and
repeated runs building up probabilities equal to the weights in the
original superposition. Enigmatic was not the entanglement of
microobject and pointer into a superposition associating each
eigenstate of the measured microobservable with a unique pointer state
(a superposition of the type often called a Schr\"odinger cat state);
such entanglement is accessible through unitary evolution, as was
already explained by von Neumann \cite{JvN}. The puzzle rather was the collapse of that superposition to the mixture with unchanged
probabilities. The current understanding is \cite{Zurek,Zeh}, that a many-freedom surrounding
decoheres the superposition. Zurek has pointed out
that decoherence of different pointer displacements is
most easily understood if the pointer displacement is
taken as the pointer's coupling agent towards a reservoir.
The prize to be paid is the distinction of a ``pointer basis'' at least for times up to $\tau_{\rm sys}$.
We now see that no such prize is due when the pointer displacement is
not the only coupling agent but just one of several, each towards a
different reservoir.

Can decoherence be reversed? Like any
other dissipative phenomenon in (subdynamics of) unitary
evolutions, decoherence should not be considered absolutely irreversible.
Seemingly spontaneous revivals of coherences out of an apparent mixture
could arise for a macroscopic system, in the unitary motion of its
composition with an environment, given a suitable initial state of the
composition. Likewise, the time reversal of decoherence is not in conflict
with unitarity of the composite dynamics and again requires ``no more
than'' suitable initial conditions. As regards
ordinary damping, such reversals were demonstrated
in the historic spin echo experiments
\cite{spinecho}; but decoherence is just ordinary damping,
starting from an extra-ordinary initial state.

While decoherence can explain why quantum superpositions are
alien to the macroworld, it does by no means imply that quantum behavior
never reaches out to the macroscopic. 
For instance, once the initial state of
the object-pointer compound of a measurement process has decohered, the pointer has vanishingly
small probability to jump between the various positions it could have gone
to by the previous interaction with the micro-object. The pointer will
rather keep moving classically, up to tiny fluctuations. What is left
from the entangled state of the micro-object and the pointer are finite
probabilities for various pointer displacements and these are due to the extremely nonclassical nature of the initial state. 

Our conclusions about the simultaneous action of several reservoirs will
not be invalidated when $f(t)$ and $F(t)$ begin to deviate from
their ${\cal O}(t^2)$ approximants. If we do not attempt to write out
corrections it is for lack of space and because the corrections are
system specific. The decoherence manifest in the ${\cal O}(t^2)$ terms
is universal, however, even beyond the harmonic-oscillator baths. To see
this we may glance back at our procedure from the reduced
generator (\ref{7}) to the density matrix (\ref{3},\ref{9}) and
its generalization to two baths. The ${\cal O}(t^2)$ behavior of
the exponentials in the density matrices (\ref{3},\ref{11}) could have
been gotten most easily by restricting ourselves to the initial form
$l(0)$ of the reduced generator: Indeed, the time integral
$\int_0^t\,dt'l(t')=l(0)t+\ldots$ and the subsequent thermal average of
its exponential precisely gives the ${\cal O}(t^2)$ terms in the
exponents in the final density matrices (\ref{3},\ref{11}); but
confining ourselves to $l(0)$ means neglecting even the free Hamiltonian
$H_{\rm bath}$ of the reservoirs and would thus not yield anything
different for quite different bath models like, say, collections of
anharmonic oscillators. The possibility of ascertaining coherence loss
from short-time expansions was also recognized in
Ref.~\cite{Brazilians}. At any rate, rapid decoherence of macroscopic
superpositions is obviously a universal phenomenon.

We owe a final remark to present efforts towards realizing
quantum computing. A quantum computer
would encorporate lots of quantum rather than classical
two-state elements and would therefore be a mesoscopic or macroscopic
complex.
The whole complex would seem prone to accelerated decoherence.
No computation relying on coherences seems possible during time spans
exceeding $\tau_{\rm dec}$. Hope must therefore be set on error correction codes or elements which
are sufficiently decoupled from detrimental environments. Symmetry may
be of help \cite{unserepapers,coherence}.
For a simple example, imagine a coupling agent of the form
$S=\Sigma^2$ such that $\Sigma$ is itself an observable which has a pair
of eigenvalues $\pm\sigma$. A reservoir contacted through the agent $S$
could not decohere a superposition within the subspace of
degeneracy, $c_+|\sigma\rangle+c_-|-\sigma\rangle$, even if the
eigenvalues $\pm\sigma$ were macroscopically distinct.

Discussions with Wojciech Zurek as well as support by the Sonderforschungsbereich ``Unordnung und Gro{\ss}e
Fluktuationen'' der Deutschen Forschungsgemeinschaft are gratefully
acknowledged.

\end{multicols}
\end{document}